\documentclass[9pt,conference]{IEEEtran}

\IEEEoverridecommandlockouts

\usepackage{graphicx}
\usepackage{enumitem}
{}
{}
{}
\usepackage{multirow}
\usepackage{color}
\usepackage[utf8]{inputenc}
\usepackage{fourier} 
\usepackage{array}
\usepackage{makecell}
\usepackage{lipsum}
\usepackage{titlesec}
\usepackage{booktabs,caption}
\usepackage[flushleft]{threeparttable}
\usepackage{amssymb,amsmath,amsthm,enumitem}
\usepackage{caption} 
\usepackage{hyperref}
\usepackage{hypcap} 
\usepackage{subcaption,siunitx,booktabs}
\usepackage[margin=1in]{geometry} 
\usepackage{caption}
\usepackage{makecell}
\usepackage{mathtools}
\usepackage{multirow}
\usepackage{subcaption}
\usepackage{booktabs,xcolor,siunitx}
\usepackage{pifont}
\newcommand{\xmark}{\ding{55}}%
\usepackage{enumitem}
\setlist{nolistsep,leftmargin=*}

\usepackage{colortbl}

\usepackage{hhline}
\usepackage{colortbl}

\begin{document}

\title{Predicting Post-Route Quality of Results Estimates for HLS Designs using Machine Learning}

\author{\IEEEauthorblockN{ Pingakshya Goswami and Dinesh Bhatia}
\IEEEauthorblockA{Department of Electrical Engineering
\\The University of Texas at Dallas, Richardson 75080,Texas}
Email: (pingakshya.goswami and dinesh)@utdallas.edu
}

\maketitle

\begin{abstract}

Machine learning (ML) has been widely used to improve the predictability of EDA tools. The use of CAD tools that express designs at higher levels of abstraction makes machine learning even more important to highlight the performance of various design steps. Behavioral descriptions used during the high-level synthesis (HLS) are completely technology independent making it  hard for designers to interpret how changes in the synthesis options affect the resultant circuit. FPGA design flows are completely embracing HLS based methodologies so that software engineers with almost no hardware design skills can easily use their tools. HLS tools allow design space exploration by modifying synthesis options, however, they lack accuracy in the Quality of Results (QoR) reported right after HLS. This lack of correctness results in sub-optimal designs with problems in timing closure. This paper presents a robust ML based design flow that can accurately predict post-route QoR for a given behavioral description without the need to synthesize the design. The model is an important design exploration tool where a designer can quickly view the impact on overall design quality when local and global optimization directives are changed. The proposed methodology presents two strong advantages: (i) Accurate prediction of the design quality (QoR), and (ii) complete elimination of the need to execute high-level synthesis for each design option. We predict three post route parameters, \textit{(i).} Area,   \textit{(ii).} Latency and  \textit{(iii).} Clock Period of a design just by analyzing the high level behavioral code and some intermediate representation codes. We have integrated the methodology with Xilinx HLS tools and have demonstrated accurate estimation on a variety of FPGA families. Our estimated results are within 10\% of actual computed values.

\end{abstract}

\medskip


\section{Introduction}\label{sec1}


The use of High-Level Synthesis (HLS) tools for FPGA-based flows is becoming common as it allows easy conversion of software descriptions into fairly accurate clocked hardware designs. HLS based design methodologies enable designers to focus on algorithmic variations of high-level programs to obtain the best micro-architectural trade-offs. HLS reduces the overall design time as it allows the generation of different versions of a functionally equivalent design without modifying the behavioral code. 

In traditional HLS design flows, there are three control parameters for generating varying RT-level designs\cite{schafer_review_paper}.  \textit{Local Synthesis Directives} in the form of \texttt{pragmas} are inserted as comments, \textit{Global Synthesis Directives} are the options like desired clock frequency that direct the overall global synthesis constraints, and \textit{Functional Unit Constraints} provide bounds on the number of available functional units of various types. The last parameter allows tighter control for SoC/ASIC style designs. Xilinx's Vivado HLS \cite{vivado_hls}, and Intel's HLS compiler \cite{intel_opencl} do not support the functional unit constraints. The only global synthesis option that the Xilinx Vivado HLX supports is the target synthesis frequency. 



\begin{figure}[t]
\center
\includegraphics[width=5cm]{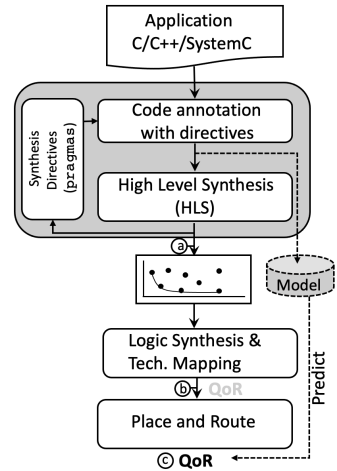}
\caption{Conventional FPGA Design Flow and the Proposed Model}
\label{basic_block} 
\vspace{-4mm}
\end{figure}

Figure \ref{basic_block} illustrates a design flow for HLS based designs targeted on FPGAs. The synthesis directives (\textit{Local} and \textit{Global} )create an annotated input for the synthesis tool. The annotated input is then synthesized along with the technology library to generate an RTL code. The process is iterated in \textit{design space exploration} many number of times to obtain an acceptable set of trade off designs (near Pareto-optimal designs). The selected optimal set of designs then undergo the steps of logic synthesis, technology mapping, and place and route.

\setlength{\belowcaptionskip}{-10pt}

\section{QoR Metrics} \label{qor_metrics}
\vspace{1mm}

HLS tools must report accurate \textit{Area} and \textit{Performance} after each synthesis run. Area reporting in the form of only the LUT usage \cite{fast_n_accurate, pyramid, ironman} and comprehensive usage of LUTs and other blocks are fairly common in the literature. In this paper, we will refer to number of LUTs ($A$) only for reporting area resources which consistent with other published works \cite{fast_n_accurate, pyramid, ironman}. Commercial HLS tools produce fairly correct estimates for the BRAM and DSP resources. 
\textit{Latency} ($L$) in the form of clock cycles is an accepted metric for FPGA-based design flows, and a correct estimate of Latency is available after the HLS Synthesis stage. In this paper, we will report Latency and the post-route maximum clock frequency $f_{max}$. 

\begin{figure}[t]
\centering
\includegraphics[width=8.2cm,  trim = 3.3cm 13.4cm 3cm 2.2cm, clip]{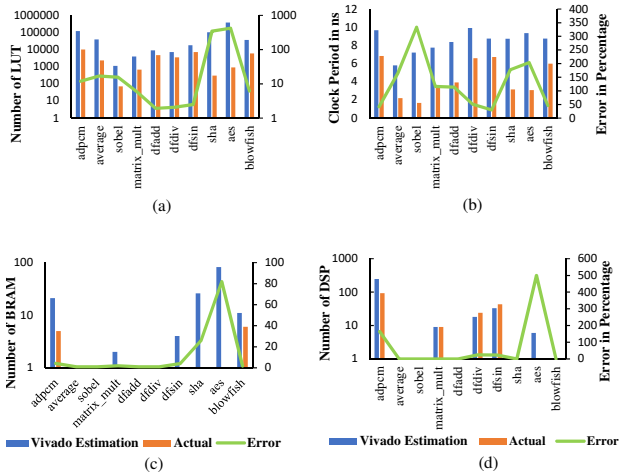}
\caption{Plots showing Vivado estimation vs actual post route values for (a) LUT Utilization; (b) Clock Period; (c) BRAM Utilization; (d) DSP Utilization}
\label{vivado_error} 
\end{figure}

The estimates produced by commercial tools for various QoR metrics after high-level synthesis at point \textcircled{a} in Figure \ref{basic_block} are usually over/underestimated by a large margin when compared to the actual Quality of Results (QoR) results produced at \textcircled{c} in Figure \ref{basic_block}. 
In our own experiments, we mapped several designs from CHStone benchmark suite\cite{chstone} using Xilinx Vivado HLS 2019.1. In Figure \ref{vivado_error}, we have plotted the actual post-route QoR values for both timing and resource requirement for the baseline designs (designs without any \textit{Local Directives}), along with the Vivado HLS error margin in percentage overlapped on the bar diagrams. Vivado HLS overestimates clock period and LUT requirement by as much as $128\%$ and $300\%$ respectively. 

\vspace{-0.1cm}
\section{Main Contribution} \label{main_contr}

\begin{table*}[t]
\centering
\caption{\small{Comparison of the proposed work with the state of the art research}}
\begin{tabular}{|l|l|l|l|l|l|l|} 
\hline
\multirow{2}{*}{\textbf{ Reference }} & \multicolumn{3}{l|}{\textbf{         Predicted Parameter }}                 & \multirow{2}{*}{\textbf{ C-Synthesis Required }} & \multirow{2}{*}{\textbf{ Feature Source }} & \multirow{2}{*}{\textbf{ Reference Labels }}  \\ 
\cline{2-4}
                                      & \textbf{ Resource } & \textbf{ Latency  } & \textbf{ Clock Period } &                                                &                                            &                                               \\ 
\hline
Fast \cite{fast_n_accurate}                         & \checkmark                   & \xmark                  & \checkmark                     & Yes                                            & Synthesis Log Files                        & Post Route                                    \\ 
\hline
Pyramid \cite{pyramid}                              & \checkmark                  & \xmark                   & \checkmark                      & Yes                                            & Synthesis Log Files                        & Post Route                                    \\ 
\hline
Comba \cite{comba_hls}                                & \xmark                  & \checkmark                  & \xmark                      & No                                             & Analytical                                 & Post C-synthesis                              \\ 
\hline
Ironman \cite{ironman}                              & \checkmark                 & \xmark                   & \checkmark                      & Scheduled DFG                                  & Scheduled DFG                              & Post Route                                    \\ 
\hline
This Work                             & \checkmark                  & \checkmark                   & \checkmark                       & No                                             & C++/LLVM IR                                & Post Route                                    \\
\hline
\end{tabular}
\label{compare_table}
\vspace{3mm}
\end{table*}

The main contribution in this paper is a robust model that takes behavioral code and constraints as an input and produces the final \textit{post-route} QoR metrics as a predicted output. The designer can use our methodology to construct a \textit{model} using the global and local features extracted from the code. The features are extracted using the LLVM compiler toolchain\cite{llvm_adve} as explained in the section \ref{mlhls}. All of the extracted features come from the LLVM Intermediate Representation (IR) frontend and some corresponding graphs.  The \textit{model} then accurately predicts post-route QoR of designs without a need to run the high-level synthesis. 
The main contributions of this paper are as follows:

\begin{enumerate}[label={$\bullet$}]
\item A methodology to extract features from high level C/C++ codes, LLVM IR codes and LLVM control/data flow and call graphs. 
\item ML based models to predict accurate and comprehensive post-route QoR from the high-level behavioral code. These include maximum frequency $f_{max}$, latency ($L$), and resources ($A$) requirement for a design, without synthesizing the designs in real time.

\item Presented results which shows the robustness of the proposed model on 10 different HLS designs and 3 different FPGA devices running at different frequencies.
\end{enumerate}

The rest of the paper is organized as follows. Section \ref{related_works_sec} gives an overview of the existing work, Section \ref{prob_def_sec} defined the problem statement. The proposed post route QoR prediction model is described in Section \ref{mlhls}. The experiments and results are presented in Section \ref{results}. Section \ref{conclusion} concludes the paper.



\section{Related Works}\label{related_works_sec}


In a recent paper, COMBA \cite{comba_hls}, the authors presented a metrics guided design space exploration method where they study the effect of different \texttt{pragmas} on the latency of the design. The research produces impressively correct performance estimates for the behavioral models in a negligible amount of time. The comparison with Xilinx Vivado HLS toolset shows that COMBA provides approximately similar quality estimates compared to the output of HLS toolset. 
However, it is observed in \cite{fast_n_accurate} as well as \cite{pyramid} that the post-synthesis QoR, as reported by commercial tools like Xilinx Vivado HLS, is grossly different from the final post-route QoR for a given design. We have highlighted this difference through experiments in section \ref{qor_metrics} as illustrated in Figure \ref{vivado_error}. Dai et al.\cite{fast_n_accurate} created a regression based machine learning model to estimate accurate post- route resource and timing of a design using features extracted from the post-synthesis log files. This work is more like calibration where resource and timing estimation error is fixed after a high-level synthesis stage. Both approaches in \cite{fast_n_accurate} and \cite{ pyramid} require time consuming C-synthesis step before QoR estimates can be corrected. In another recent work, Ironman \cite{ironman}, the authors proposed post route resource and timing requirement of HLS designs using graph based learning methods. In order to do so, they used synthesized data flow graphs(DFG) generated from IR code as an input to their ML model.

While fairly accurate, the methods reported in \cite{fast_n_accurate} and \cite{pyramid} require a time-consuming synthesis step before QoR can be accurately estimated and \cite{ironman} requires scheduled DFG graphs from synthesis stage. COMBA is a fast approach for estimation but the accuracy of estimates matches only with the post C-synthesis results. In Table \ref{compare_table}, we have summarized the four relevant works described in \cite{fast_n_accurate, pyramid, ironman} and \cite{comba_hls}, and compared them with our presented research. The research presented in this paper is comprehensive and predicts all the post route vital performance measuring parameters of a HLS designs viz., clock period, latency and resource requirement together. It also eliminates the need for full C-synthesis.


\vspace{1mm}

\section{Problem Definition}\label{prob_def_sec}
\vspace{2mm}



Given a synthesizable C/C++ based high-level code with synthesis directives in the form of \texttt{ pragma}, and the desired global frequency of operation, create a machine learning based model which will predict the post-route clock period, latency and resource requirement of a design without synthesizing the design in real time.

We formulated the post route QoR prediction of HLS design as a machine learning based regression model. Fig. \ref{basic_block} shows the basic block diagram of our proposed methodology. 




\section{QoR Metrics Prediction Framework}\label{mlhls}

LLVM\cite{llvm_adve} is a open source compiler which is used to develop executable codes for multiple platforms. LLVM generates technology and processor independent assembly language code called \textit{Intermediate Representation (IR)}. There is an optimizer present in LLVM, which optimizes the IR code. The final optimized IR code is converted into Control and Dataflow Graphs (CDFGs) which are later used for scheduling and binding of the generated RTL code. The LLVM toolchain also generates callgraphs which represent the inter-function relationships in a C/C++ codes. These functions are later synthesized as individual modules in the generated RTL code.

We have analyzed the High Level C/C++ codes, \texttt{Pragma} optimized LLVM IR codes, Control and Dataflow Graphs, and Call-graphs to extract features that can help us in predicting final QoR metrics. We have used HLS benchmarks taken from  CHStone\cite{chstone} benchmark set. These benchmarks are very diverse in terms of code structure, applications as well as area and timing costs. These benchmarks represent many single-function and multi-function designs. We generate multiple versions from a single HLS design by changing the following six \texttt{pragmas} in the HLS codes: (i) Loop Unrolling, (ii) Loop Pipelining, (iii) Array Partitioning, (iv) Array Reshaping, (v) Function Inlining, and (vi) Function Instantiation. 

For each benchmark shown in Table \ref{train_table}, we have generated 400 versions (data) of the design by varying the \texttt{pragmas} and desired \textit{target frequency}. Table \ref{train_table} shows the statistics of range of variation of Latency, Clock-Period, and the LUT usage for each benchmark versions when the designs are mapped on the Zynq 7000 device. Our methodology has also been verified using Xilinx VIRTEX and KINTEX series of devices also. It is clear from Table \ref{train_table} that the data is very diverse and would support building robust training models.

\vspace{+2mm}
\begin{table}[h]
\caption{Statistics of Training/Testing benchmarks}
\centering
\begin{tabular}{p{1.5cm} p{1.5cm} p{1.5cm} p{1.5cm}}
\hline
\begin{tabular}[c]{@{}l@{}}\textbf{Design }\\\textbf{~Name}\end{tabular} & \begin{tabular}[c]{@{}l@{}}\textbf{Latency }\\\textbf{~Range}\end{tabular} & \begin{tabular}[c]{@{}l@{}}\textbf{Clock Period}\\\textbf{~Range (ns) }\end{tabular} & \textbf{ LUT Range }  \\ 
\hline\hline
adpcm                  & 3119-62536                                                                 & 2.078-8.482                                                                          & 56-410000              \\ 
\hline
matrix\_mult           & 38-274                                                                     & 3.48-6.849                                                                           & 667-7139              \\ 
\hline
sobel                  & 15-38                                                                      & 1.463-8.44                                                                           & 33-121                \\ 
\hline
average                & 2-211                                                                      & 1.338-2.623                                                                          & 4-3265                \\ 
\hline
dfadd                  & NA                                                                         & 3.908-6.616                                                                          & 3634-4891             \\ 
\hline
dfsin                  & NA                                                                         & 6.13-8.74                                                                            & 4348-8818             \\ 
\hline
dfdiv                  & NA                                                                         & 6.687                                                                                & 3318-3472             \\ 
\hline
sha                    & 45-1330                                                                    & 2.209-4.221                                                                          & 101-1889              \\ 
\hline
blowfish               & 23525-36386                                                                & 4.421-9.371                                                                          & 3357-60537            \\ 
\hline
aes                    & 2697-2937                                                                  & 2.615-5.021                                                                          & 166-1639              \\ 
\hline\hline
\textbf{Overall}                & 2-63536                                                                    & 1.436-9.371                                                                          & 4-60537               \\
\hline
\end{tabular}
\label{train_table}
\end{table}

\vspace{-0.2cm}



\subsection{Feature Extraction}\label{feature_ext}

We have created an extensive set of features to create a robust prediction model. We extracted a total of \textbf{69} features from four different sources, examples of which are shown in Table \ref{feature_table}.  The features are either \textit{numerical} or \textit{categorical} features. An example of a \textit{numerical} feature is the length of the longest path in the data-flow graph, and an example of a \textit{categorical} feature is data-type that can be 8-bit integer, 16-bit integer, or a double-precision floating point number. Table \ref{feature_table} lists some of the important features used for training the models. The extracted features from each of the four sources are briefly described here.


\begin{table}[b]
\centering
\caption{\small{List of Features from each source.}}
\begin{tabular}{|p{1cm}|p{4.5cm}|p{1.3cm}|} 
\hline
\textbf{Feature Source } & \textbf{Feature Examples }  & \textbf{No. of \newline Features }  \\ 
\hline
HLS Code                  & \begin{tabular}{@{\labelitemi\hspace{\dimexpr\labelsep+0.5\tabcolsep}}l@{}} Max and average unrolled factor\\Max \& Average Batch size of loops\\Max Pipelined Loop Name\\Max \& Average Pipelined II ~ \end{tabular} & 13                             \\ 
\hline
LLVM IR Code              &
\begin{tabular}{@{\labelitemi\hspace{\dimexpr\labelsep+0.5\tabcolsep}}l@{}} Max, average and total number of \\ instructions per BB\\Max, average, total number of  \end{tabular} (math, sign ext, zero ext, logic, memory, vector, other) operations
& 44                             \\ 

                  
\hline
CDFG                      &

\begin{tabular}{@{\labelitemi\hspace{\dimexpr\labelsep+0.5\tabcolsep}}l@{}} Total number of nodes \\ Max length of critical path \\Number of FCUs \\Max path length\\Max and average number of  \end{tabular} incoming/outgoing edges & 6                              \\ 

\hline
Callgraph                 & \begin{tabular}[c]{@{}l@{}}\begin{tabular}{@{\labelitemi\hspace{\dimexpr\labelsep+0.5\tabcolsep}}l@{}} Properties of child functions\end{tabular}\\Max, min (latency, CP, FCUs)\end{tabular}                                                                                & 6                              \\ 
\hline
Total                     & ~                                                                                                                                                                                                                                                                           & 69                             \\
\hline
\end{tabular}
\label{feature_table}
\end{table}

\begin{enumerate}[leftmargin=*,label=\roman*., itemsep=0pt, topsep=0pt, wide,leftmargin=0cm]
    \item \textbf{\textit{High Level Code}}: As shown in Table \ref{feature_table}, we extract 13 features from the HLS code. These features include information about the \texttt{pragmas}. Features extracted from HLS code include information about the loop unrolling and loop pipelining. We extract unroll factor, pipeline initiation interval (II) and batch size of the loops. Based on the  \texttt{pragma} arguments, we compute a \texttt{batch\_size} for each loop. It is defined as a ratio between the loop-bound and the unrolling factor for a loop. While the unrolling factor represents the degree of parallelism, and the \texttt{batch\_size} represents the depth of the computation.

  \item \textbf{\textit{LLVM IR Code}}: LLVM IR code are made up of basic blocks (BB). Each BB is a set of continuous assembly instructions, which are branched at if-else conditions, function returns and loops.  We analyzed each of the BB in detail, and calculated total, maximum and average number of instructions in each BB. 
  There are 9 types \cite{llvm_instructions} of instructions in LLVM. From each BB, we calculate the maximum, total and average number math, sign/zero extension, memory and logic instruction and vector instructions   from the LLVM IR code. Example features are shown in Table \ref{feature_table}. This is most exhaustive source of features, and we extract 44 featues from the IR code. 
  
  \item \textbf{Control and Data Flow Graph}: Control flow graphs \texttt{CFGs} show how the basic blocks (BB) communicate with each other,  propagation of instructions and inter BB dependencies. Data flow graphs (\texttt{DFGs}) show flow of variables between BBs. \texttt{DFGss} are used to avoid Read-After-Write and Write-After-Read hazards.  Features like total number of BBs, number of BBs in the longest path, number of incoming and outgoing edges on a BB, data types of the edges connecting the BB are extracted from the CDFG. Pre-synthesis functional unit (FCU) counts based on LLVM passes is another feature extracted from CDFGs. The features from CDFGs directly effect the latency of a design. 
  
  \item \textbf{Callgraph}:  \textbf{\textit{Callgraphs}} describe the inter-function relationship in a code. Most of the HLS codes are complex multi\_function designs. Each node in the callgraph represents a function in the HLS code. By analyzing the callgraph, characteristics of the child nodes like a number of FCUs, latency, minimum and maximum clock period of child functions and the datatype passed from one function to another are extracted and passed to the root (Top) function. 

\end{enumerate}

\begin{figure}[h]
\center
\includegraphics[width=7cm,  trim = 3cm 19.2cm 8cm 2.8cm, clip]{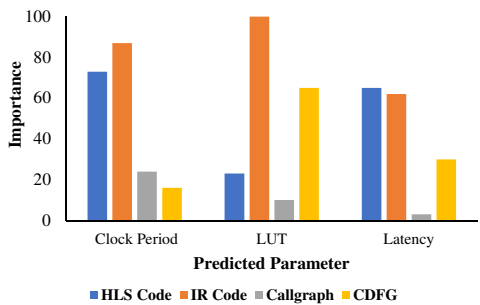}
\caption{Feature Importance Analysis}
\label{feature_imp} 
\end{figure}

We passed all the features through a feature analysis tool, \texttt{xgboost} regression feature analysis tool \cite{xgb_feature} and plotted the results in Fig. \ref{feature_imp}. In Fig. \ref{feature_imp}, we have \textit{normalized} the importance of each feature to a range of 0 to 100. Each bar in Fig. \ref{feature_imp} represents the sum of all the features from a particular source.  As it can be seen from the bar diagram, IR file features hold most importance, while callgraph features hold least importance for our regression model.

\medskip

\begin{figure}[t!]
\center
\includegraphics[width=7cm,  trim = 0cm 18.6cm 10cm 0.8cm, clip]{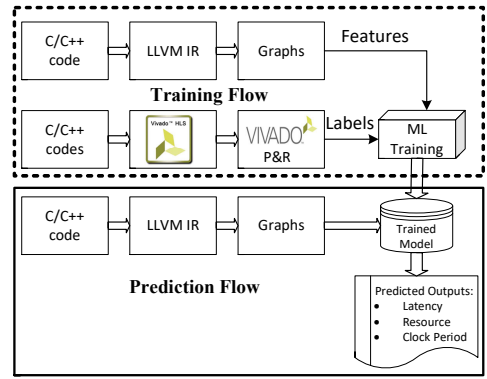}
\caption{\small Training and Prediction Flow}
\label{training_flow} 
\vspace{0.4cm}
\end{figure}

\subsection{Model Creation} \label{model_creation}

  \begin{figure}[b]
\center
\includegraphics[scale=0.55,  trim = 1cm 16.2cm 5cm 2.4cm, clip]{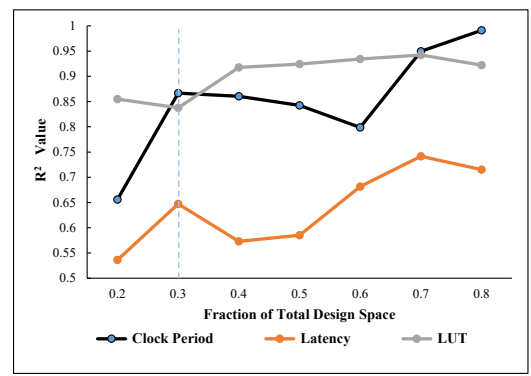}
\caption{Change in $R^2$ value with increase in training datasize}
\label{acc_data} 

\end{figure}

For each data element described in the section \ref{mlhls}, we extract features. The same designs are also routed using Vivado P \& R tools to generate the target labels. We also observed that the labels can be generated at after logic synthesis stage also, step \textcircled{b} in Figure \ref{basic_block} when the accurate estimate of resources and a near-accurate estimate of \textit{clock frequency} and \textit{latency} is available to the designer. There is a small error for clock period, since the router adds the wire delay thereby increasing the post route clock period by a small amount- around 1.5 ns. Labels generated after logic synthesis will also provide same QoR but we will save substantial (almost 30\%) place \& route execution time for label creation.
For this particular work, we used the labels generated after routing. The entire training and prediction flow is illustrated in Fig. \ref{training_flow}, where the black dotted box represents the training part consisting of feature extraction, label generation, and training, and the solid box below represents the prediction flow.

In order to measure the correctness of QoR prediction, we have used the \textit{average of mean absolute percentage error} (MAPE) for all the models. MAPE is calculated using the following formula:

\begin{equation}
MAPE=  \frac{1}{N}\sum\left|{\frac{y_{actual}-y_{predicted}}{y_{actual}}}\right|*100
\label{eq1}
\end{equation}

In equation \ref{eq1}, $y_{predicted}$ and $y_{actual}$ represents the predicted and actual value respectively while N represented the total number of designs being tested.

\section{Testing and Results}\label{results}
\medskip
\subsection{Experimental Setup}

We have synthesized and implemented our HLS designs on Intel Xeon E5-2603 quadcore processor based workstation with 32 GB RAM. The synthesis and implementation software used for this work is Xilinx Vivado 2019.1, while the target hardware is Xilinx Zynq 7000 series FPGA device. Our methodology has also been verified using Xilinx VIRTEX and KINTEX series of devices and we will present to support the versatility of our methodology. 
We have used state of the art ML libraries ScikitLearn 0.19.1\cite{scikit} and xgboost 0.90 \cite{xgb_feature} running on  Nvdia Geforce GTX 1080 GPUs to train our models. We used cloud-based Comet.ml \cite{comet_ml} for hyperparameter tuning tool. 


\subsection{Training of Models}

\begin{table}[t]
\centering
\caption{\textbf{MAPE} for different models trained}
\begin{tabular}{|p{1.9cm}|p{1.4cm}|p{1.4cm}|p{1.4cm}|} 
\hline
\multicolumn{4}{|c|}{\textbf{Model Comparison}}  \\ 
\hline
\textbf{Model~ Used} & \textbf{CLK} & \textbf{Latency} & \textbf{LUT}             \\ 
\hline
HLS                  & 100.45\%       & NA               & 392.33\%                    \\ 
\hline
MLP                  & 9.69\%      & 16.58\%          & 44.14\%                  \\ 
\hline
RF                   & 7.98\%      & 18.25\%          & 16.28\%                  \\ 
\hline
{\bf XGB}                  & {\bf 6.29\%}      & {\bf 10.22\%}           & {\bf 10.32\%}                  \\
\hline
\end{tabular}
\vspace{-3mm}
\label{qor_table}
\end{table}

 \begin{figure*}[t]
\center
\includegraphics[width=14cm,  trim = 3.0cm 3.6cm 3.0cm 3.7cm, clip]{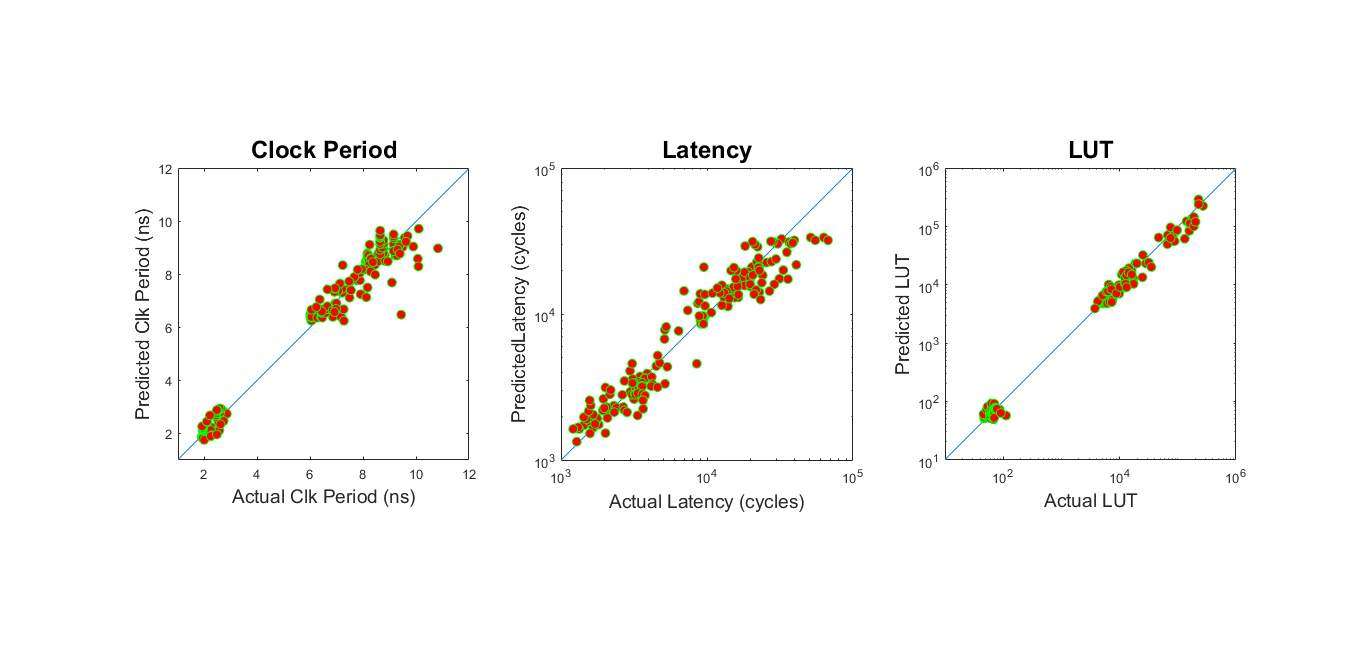}
\caption{Actual vs Predicted Values for the timing, latency and resource for ADPCM  on Zynq 7000 (100MHz)}
\label{acc_vs_predicted_all} 
\vspace{0.3cm}
\end{figure*}


 As stated in section \ref{qor_metrics}, we have used 400 versions of designs for each of the benchmarks. To study the effectiveness of a regression  model, we have plotted how the value of $R^2$ increase with increase in training data size in Figure \ref{acc_data}. The x-axis represent the percentage of total design points used for training. 
 An $R^2$ value in the range of $0.7$ to $0.8$ is considered to be good for a model. The plot in the Figure \ref{acc_data} shows that a well-trained model can be achieved using a small fraction (around 30\%) of the total data for predicting LUT resources and the clock period. However, predicting for \textit{latency} may require much larger training data set. Based on assertions presented in Figure \ref{acc_data}, we trained the individual designs on the generated 120 versions and predicted on another 280 designs. The \textit{matrix\_multiplier} design was trained and tested on half of the total 1500 designs.  This is because  \textit{matrix\_multiplier} is highly customizable in terms of dimensions, \texttt{pragmas} and frequencies.

\subsection{Analysis of Results}

We have presented results using all of the ten benchmarks described in Table \ref{train_table}. We have also conducted experiments to demonstrate the robustness of our models using one very large and diverse benchmark: \texttt{adpcm}. \texttt{adpcm} is a very complex benchmark consisting of 13 loops, 10 arrays and 11 different functions.  We selected \texttt{adpcm} because it is the most diverse design in terms of resource and latency requirement among all the CHStone benchmarks. As shown in Table \ref{train_table},the LUT utilization ranges from 56 to 410K LUTs, while flipflop utilization ranges from 83 to 237k units and latency ranges from 3119 to 62536 clock cycles.


\medskip

\begin{table*}[t]
\centering
\caption{ Actual vs Predicted values of resource and timing. Frequency Sweep from 100MHz t0 500 MHz }
\begin{tabular}{|l|l|l|l|l|l|l|l|l|} 
\hline
~                         & \multicolumn{3}{l|}{\textbf{ Clock Period (ns) }}                & \multicolumn{2}{l|}{\textbf{ Latency (clock cycles) }} & \multicolumn{3}{l|}{\textbf{ \# of LUTs }}                               \\ 
\cline{2-9}
\textbf{ Target Frequency(MHz) } & \textbf{ Actual } & \textbf{ Predicted } & \textbf{ Vivado HLS } & \textbf{ Actual } & \textbf{ Predicted }               & \textbf{ Actual } & \textbf{ Predicted } & \textbf{ Vivado HLS }  \\ 
\hline
100                       & 7.74              & 7.44                 & 9.10                  & 20104             & 21659                              & 2782              & 3244                 & 5038                   \\ 
\hline
125                       & 6.64              & 7.41                 & 8.30                  & 23604             & 21677                              & 2774              & 3244                 & 5039                   \\ 
\hline
150                       & 6.64              & 6.29                 & 6.05                  & 28254             & 26916                              & 2772              & 3244                 & 5852                   \\ 
\hline
175                       & 4.55              & 4.49                 & 5.55                  & 34004             & 40351                              & 2722              & 2605                 & 5821                   \\ 
\hline
200                       & 4.55              & 4.60                 & 4.49                  & 43154             & 41332                              & 2588              & 2735                 & 4487                   \\ 
\hline
225                       & 4.55              & 4.60                 & 3.83                  & 43204             & 41332                              & 2636              & 2735                 & 4484                   \\
\hline
300                       & 3.00              & 3.11                 & 3.39                 & 55954             & 43586                              & 2740              & 2749                 & 5030                   \\
\hline
500                       & 3.00              & 3.46                & 3.39                 & 111654             & 42881                              & 3343              & 5095                 & 6687
                   \\

\hline
\end{tabular}
\vspace{+5mm}
\label{freq_experiment}
\end{table*}

\begin{table*}[h]
\centering
\caption{Validation and Test \textbf{MAPE} on 64 unseen designs on 3 different FPGA devices}
\begin{tabular}{|l||l|l|l||l|l||l|l|l|} 
\hline
~                 & \multicolumn{3}{c||}{\textbf{ Clock Period }}                    & \multicolumn{2}{c||}{\textbf{ Latency }} & \multicolumn{3}{c|}{\textbf{ LUT }}                              \\ 
\cline{2-9}
\textbf{ Device } & \textbf{ Validation } & \textbf{ Test } & \textbf{ Vivado HLS } & \textbf{ Validation } & \textbf{ Test } & \textbf{ Validation } & \textbf{ Test } & \textbf{ Vivado HLS }  \\ 
\hline
Zynq 7000         & 5.55                  & 7.75            & 198.09                & 20.24                 & 17.54           & 19.02                 & 18.94           & 529.27                 \\ 
\hline
Virtex 7          & 4.11                  & 6.50            & 152.66                & 18.36                 & 17.10           & 12.36                 & 17.89           & 362.52                 \\ 
\hline
Kintex 7          & 5.51                  & 5.06            & 82.6                  & 17.69                 & 19.50           & 14.14                 & 11.70           & 531.29                 \\
\hline
\end{tabular}

\label{tab:different_devices}

\end{table*}

\begin{table}[h]
\centering
\caption{ \small{\textbf{MAPE} for benchmarks and Vivado  Estimates}}
\begin{tabular}{|p{1.0cm}||p{0.7cm}|p{0.8cm}||p{1cm}||p{0.8cm}|p{0.8cm}|} 
\hline
\textbf{~}                              & \multicolumn{2}{l|}{\textbf{Clock Period}}                                     & \textbf{Latency}                      & \multicolumn{2}{l|}{\textbf{Resource}}                                           \\ 
\hline
\textbf{Design Name\textcolor{white}{}} & \textbf{This Work\textcolor{white}{}} & \textbf{Vivado HLS\textcolor{white}{}} & \textbf{This Work\textcolor{white}{}} & \textbf{This Work \textcolor{white}{}} & \textbf{Vivado HLS\textcolor{white}{}}  \\ 
\hline
adpcm                                   & 7.75                                  & 198.09                                 & 17.54                                 & 18.94                                  & 529.27                                  \\ 
\hline
ave8                                    & 7.28                                  & 182.14                                 & 16.41                                 & 7.82                                   & 73.18                                   \\ 
\hline
matmul                                  & 7.92                                  & 67.15                                  & 11.12                                 & 8.37                                   & 560.78                                  \\ 
\hline
sobel                                   & 0.6                                   & 163.23                                 & 1.76                                  & 0.94                                   & 584.47                                  \\ 
\hline
dfadd                                   & 6.33                                  & 87.88                                  & 4.3                                   & 2.04                                   & 98.58                                   \\ 
\hline
dfdiv                                   & 0.19                                  & 26.19                                  & NA                                    & 0.03                                   & 114.38                                  \\ 
\hline
dfsin                                   & 2.78                                  & 39.39                                  & NA                                    & 3.78                                   & 167.21                                  \\ 
\hline
aes                                     & 9.44                                  & 103.46                                 & NA                                    & 20.78                                  & 911.25                                  \\ 
\hline
blowfish                                & 14.35                                 & 36.57                                  & NA                                    & 30.25                                  & 491.91                                  \\ 
\hline
\textbf{Average}                        & \textbf{6.29}                         & \textbf{100.45}                        & \textbf{10.22}                        & \textbf{10.32}                         & \textbf{392.33}                         \\
\hline
\end{tabular}
\label{individual_results}
\begin{tablenotes}
      \small
      \item \textbf{Note:} \small{NA: Reference labels not reported by HLS tool}
    \end{tablenotes}
\end{table}

We have trained our model using three different regression models viz. \textit{(i)} Multi-layer Perceptron Regression (MLP);  \textit{(ii)}  Random Forest (RF)   \textit{(iii)} Gradient Boost Regression (XGB). The average results for the 10 benchmarks using the three models are shown in Table \ref{qor_table}. In Table \ref{qor_table}, we have shown the average MAPE for the clock, latency, and LUTs for the ten benchmarks using the three regression models. In the third row (HLS), average MAPE reported by Vivado HLS after C-synthesis stage is presented. Each error is computed against the actual post-route value. It can be observed that very high accuracy is achieved using ensemble methods like XGB and RF. We chose XGB as our prefered model for rest of the experiments.

Table \ref{individual_results} shows the MAPE values for each of the ten benchmarks used for testing. These results are based on XGB regression model, where training is done on $120$ versions of a design, and prediction is done on $280$ versions. Vivado HLS estimates are shown in Column 3 and 6 in Table \ref{individual_results}. The estimates produced by Vivado HLS are very inaccurate, and on an average, our prediction is very accurate and our framework promises to provide a good design guidance to the designer. The LUT resource usage error change is relatively higher when compared with the clock period and the latency of the design. This is because the spectrum of LUT usage is very wide from a minimum of 4 to a maximum of 65K LUTS. In Fig. \ref{acc_vs_predicted_all}, we have show the actual vs predicted values of all the three predicted parameters for \texttt{adpcm} benchmark, running on Zynq 7000 FPGA device using 100MHz frequency. As it can be seen from the figure, our prediction highly correlates with the actual value. 


Our model is able to capture the effect of the target \textit{global frequency} on the overall design and resource usage. It is expected that the variation of desired global frequency will result in a change in \textit{scheduling} during the high-level synthesis, and therefore will result in different architectures with varying resource usage, latency, and clock-period. We tested the model for \texttt{adpcm} benchmark. We tested the model on the \textit{baseline} (no \texttt{pragmas} design for \texttt{adpcm}). During training, we generated labels using three target frequency selections, i.e., $100 MHz$, $150 MHz$, and $200 MHz$. During testing, we varied the global-desired-frequency constraint from $100 MHz$ to $500MHz$. After $500 MHz$, Xilinx Vivado returns the same best possible design. Normally, an HLS tool will schedule single computation step in multiple clock cycles when the desired target frequency is very high. Xilinx Vivado does not perform such mapping. Table \ref{freq_experiment} illustrates the results. We can observe that our model results in very accurate estimates for the QoR metrics. 
It is worth mentioning there that we trained on 3 frequencies ($100MHz$, $150MHz$ and $200MHz$), while the model is able to accurately predict on 8 frequencies, 5 of which are not part of the training dataset. In Table \ref{freq_experiment}, we have also shown the estimated values of Vivado HLS tool after C-synthesis.


 We also demonstrate the robustness of our methodology on other devices from Xilinx, including the Virtex-7 and Kintex-7 series of FPGAs. We trained the models using labels when designs were mapped on the selected device. Table \ref{tab:different_devices} presents results when \texttt{adpcm} benchmark was mapped on Virtex, Kintex, Zynq devices.  The \textit{test} results in Table \ref{tab:different_devices}, however, are based on 64 completely new (blind) versions of \texttt{pragma} inserted designs that were not part of the train and validation framework. 

\medskip

\section{Conclusion}\label{conclusion}

We have presented a comprehensive and robust classical machine learning based post-route timing and resource requirement prediction tool for FPGA based HLS designs. Unlike previous approaches \cite{fast_n_accurate, pyramid, ironman}, our methodology does not require time consuming C-synthesis step as we extract features right out of the high-level behavioral code. Our method improves the state of the art by producing better or comparable results using a model created prior to synthesis.
 We believe that this is first such attempt that models the entire design flow as a machine learning model, thereby totally eliminating the HLS and physical design tools from the flow.

\nocite{*} 

\end{document}